\def\be{\begin{equation}}
\def\ee{\end{equation}}
\def\ber{\begin{eqnarray}}
\def\eer{\end{eqnarray}}
\def\bers{\begin{eqnarray*}}
\def\eers{\end{eqnarray*}}

\documentclass[aps,prb,twocolumn,showpacs,groupedaddress,amsmath,amssymb]{revtex4-1}
\usepackage{dcolumn}   
\usepackage{amsmath}
\usepackage{titlesec}
\titlespacing*{\section}{0pt}{0.8\baselineskip}{0.6\baselineskip}
\usepackage[colorlinks=true,citecolor=black,linkcolor=blue,urlcolor=blue]{hyperref}
\usepackage{multirow}
\usepackage{graphicx}    
\usepackage{subfigure}
\usepackage{romannum}
\usepackage{comment}
\usepackage{color}
\usepackage{makecell}
\usepackage{ifthen}
\newboolean{includefigs}
\setboolean{includefigs}{true}      
\newboolean{includetext}
\setboolean{includetext}{true}     
\newcommand{\condcomment}[2]{\ifthenelse{#1}{#2}{}}

\begin{document}

\title{Cu$_2$ZnSiTe$_4$: A potential thermoelectric material with promising electronic transport}
\author{Himanshu Sharma$^1$, Bhawna Sahni$^1$, Tanusri Saha-Dasgupta$^2$ and Aftab Alam$^1$}
\email{aftab@iitb.ac.in}
\affiliation{$^1$Department of Physics, Indian Institute of Technology, Bombay, Powai, Mumbai 400 076, India}

\affiliation{$^{2}$Department of Condensed Matter and Materials Physics, S. N. Bose National Centre for Basic Sciences, JD Block, Sector III, Salt Lake, Kolkata, West Bengal 700106, India}

\begin{abstract}
Transition metal-based quaternary chalcogenides have gathered immense attention for various renewable energy applications including thermoelectrics (TE). While low-symmetry and complex
structure help to achieve low thermal conductivity, the TE power factor and hence the figure of merit (ZT) remains low which hinders to promote these class of materials for future TE applications. Here, we investigated the TE properties of a new system, Cu$_2$ZnSiTe$_4$, with improved electronic transport using first-principles calculation. The presence of heavy chalcogen like Te, helps to achieve a relatively low bandgap (0.58 eV). This, together with unique electronic band topology, leads to a promising value of power-factor of 3.95(n-type) and 3.06(p-type) mWm$^{-1}$K$^{-2}$ at 900 K. Te atoms also play a crucial role in mixing the optical and acoustic phonon branches which, in turn, are responsible for reduced lattice thermal conductivity ($\sim$0.7 Wm$^{-1}$K$^{-1}$ at high temperature). Though the thermal conductivity is not appreciably low, the electronic transport properties (power factor) are quite favorable to yield promising TE figure of merit (ZT $\sim$2.67 (n-type) and $\sim$2.11 (p-type) at 900 K). We propose Cu$_2$ZnSiTe$_4$ to be a potential candidate for TE applications, and believe to attract future experimental/theoretical studies.
\end{abstract}
\date{\today}

\maketitle
\section{Introduction}

The limitation of fossil fuel reservoirs and their adverse effects on the environment demands alternate ways of generating renewable energy. Although, there exist several methods of producing renewable energies such as solar, wind or hydro-power etc., yet a large fraction of energy gets wasted in the form of heat while using primary energy resources. Technology based on thermoelectric (TE) devices can help in converting the waste heat into electricity in an eco-friendly manner. Durability over time, zero pollution and no toxic outputs make such devices more useful for renewable energy. These devices require specific types of materials which dictates their efficiency via a thermoelectric figure of merit, defined as ZT=$\frac{S^2\sigma}{\kappa_e + \kappa_l}$, where S, $\sigma$, $\kappa_e$ and $\kappa_l$ are the  Seebeck coefficient, electrical conductivity , electronic and lattice contribution to the thermal conductivity. The inter-coupling of these physical quantities for a given TE material  makes it challenging to achieve higher conversion efficiency, and hence the hunt for optimal candidate material remain an ongoing effort in the TE community.

 One of the ways to improve efficiency is to optimize the carrier concentration to enhance the power-factor($S^2\sigma$) and hence the figure of merit. Compounds with ultra-low lattice thermal conductivity ($\kappa_l$) is yet another strategy to achieve high ZT. The search for materials with such properties has led researchers to a wide variety of materials like clathrates\cite{clathartes1}, skutterudites\cite{skutterdite}, half-heuslers\cite{half-heusler} and chalcogenides.
In particular, chalcogenides have a variety of interesting properties which make them promising for TE applications. Among these, quaternary chalcogenides\cite{book,review_arc1}, especially copper(Cu)-based, have acquired tremendous popularity in the past few years stimulated by their use in several energy related research such as solar cell, magnetoelectric, photocatalysis etc.\cite{Solarcel_qe,magnet_qe2,photo_qe} In most cases, Cu-based quaternary chalcogenides, with a general formula Cu$_2$ABX$_4$, where A is a transition metal, B is group IVA element like Si/Ge/Sn and X is chalcogen, crystallize in a tetragonal structure with space group I$\bar{4}$2m. The relatively low structural symmetry compare to parent binary compounds and presence of the chalcogen (S, Se, Te) atoms, are known to facilitate low lattice thermal conductivity, thus being attractive for possible TE applications. Quaternary chalcogenides show the lattice thermal conductivity of the order of around 2-3 W/m-K \cite{p3py1, Co_Mn3, HgSn1, ZnSn1} where as binary compounds (ZnS/Se/Te)\cite{binary1, binary2} shows it around 20 w/m-K. Quaternary chalcogenides are also interesting because of the distinct electronic features exhibited by their two structural units (1) the copper-chalcogenide unit being responsible for the electrical transport and (2) ABX$_4$ unit which remains insulating. Thus, quaternary chalcogenides are potential candidates to optimize the electronic and thermal transport simultaneously and hence achieve higher TE efficiency.

However, most of the reported quaternary chalcogenides are found to show wide band gaps ($\sim$ 1 eV), being a hindrance for promising electrical conductivity. Even some of the most studied materials in this family such as Cu$_{2}$ASnSe$_{4}$ where A=Co,Mn,Cd,Zn show conductivity of the order of $10^3$-$10^4$ Sm$^{-1}$ lower than their binary counterparts, PbTe and SiGe ($\ge 10^4$ Sm$^{-1}$)\cite{Co_Mn3,CdSn2,ZnSn1}. Additionally, the asymmetry associated with structure reduces band degeneracy around the Fermi level and hence degrades the Seebeck coefficient. These result into a moderate value of ZT ($\sim$0.14-0.7)\cite{CdSn2,Cd_doped1,Co_Mn3}. Among these compounds, Cu$_2$CoSnSe$_4$ (with ZT = 0.7) shows a power-factor of 0.67 mWm$^{-1}$K$^{-2}$ at the intrinsic carrier concentration. Other compounds, e.g. Cu$_2$(Zn, Cd, Mn)SnSe$_4$, show even smaller power factor lying in the range 0.2-0.38 mWm$^{-1}$K$^{-2}$  at $\sim$800 K.
 It has been suggested that non-stoichiometry, such as Cu excess at A/B site in Cu$_2$ABX$_4$, can increase the carrier concentration and hence the electrical conductivity, leading to better power-factor values\cite{review2}. For example, Cu excess at Mn site in Cu$_2$MnSnSe$_4$\cite{Mn_doped1} increases $S^2\sigma$ from 0.38 to 0.7 mWm$^{-1}$K$^{-2}$ and at Cd site in Cu$_2$CdSnSe$_4$\cite{Cd_doped1} increases it from 0.16 to 0.5 mWm$^{-1}$K$^{-2}$ resulting in ZT values of 0.61 and 0.65 respectively. Although variation in stoichiometry helps to increase conductivity, the overall enhancement in TE performance is not appreciable. This is partly due to the fact that though this strategy increases $\sigma$, it concurrently reduces the Seebeck coefficient and increases the electronic thermal conductivity. Thus, it is very essential to seek alternative means to enhance the electrical conductivity, together with optimizing the carrier concentration in such a way to achieve a reasonable Seebeck coefficient and thermal conductivity in these family of compounds.

Here, we propose Cu$_2$ZnSiTe$_4$ to be a promising candidate in this family for both n-type and p-type conduction. This compound has been synthesized earlier and characterized through X-ray and Raman scattering probes.\cite{RamanCu2Te4} However, to the best of our knowledge, its electronic and thermal transport properties are never studied, specially from a TE point of view. In this study, we carry out a detailed first principles calculation to evaluate the same. First of all, we show that the inclusion of Tellurium reduces the band gap due to its significant spin-orbit coupling (SOC) effect, enhancing the electrical conductivity. SOC, however, adversely affects the degeneracy of the bands\cite{degen1,degen2} as well, particularly near the valence band, lowering the Seebeck coefficient. As such, a careful optimization of carrier concentration is needed to achieve the optimal value of power factor. To get an estimate of the optimal carrier concentration, we have simulated the TE coefficients in the concentration range 10$^{18}$-10$^{20}$ cm$^{-3}$ between 300 and 900 K. A power factor value of S$^2\sigma$=3.95 mWm$^{-1}$K$^{-2}$ is found at 10$^{19}$ cm$^{-3}$ for n-type and S$^2\sigma$=3.28 mWm$^{-1}$K$^{-2}$ at 5$\times$10$^{19}$ cm$^{-3}$ for p-type conduction. Interestingly Te, being a heavy element, also enhances the coupling between optical and acoustic phonon branches causing a reduction in the lattice thermal conductivity. This eventually leads to a high ZT value of 2.67 for n-type and 2.11 for p-type, at 900 K. Thus, Cu$_2$ZnSiTe$_4$ offers
promising electronic transport with relatively low lattice thermal conductivity. The present study is expected to motivate future experimental research on this compound as a promising
candidate for TE applications.

\section{Computational Details}
First-principles calculations were performed using density functional theory (DFT) as implemented in Vienna Ab initio Simulation Package (VASP) with the projector augmented wave(PAW) method.\cite{vasp1, vasp2, vasp3} Generalized gradient approximation (GGA) exchange-correlation functional parametrized by Perdew-Burke-Ernzerhof (PBE)\cite{GGA}  was used to calculate the electronic structure including spin-orbit coupling (SOC) effect. To make an accurate estimate of the band gap, the modified Becke-Johnson (mBJ)\cite{mbJ} meta-GGA exchange-correlation functional including SOC was employed. A plane wave energy cut-off of $500$ eV is used with $\Gamma$-centred k-mesh to sample the entire Brillouin zone (BZ), with an energy convergence of $10^{-6}$ eV. We used density functional perturbation theory (DFPT)\cite{dfpt} as implemented within VASP for phonon dispersion calculation.\cite{phonopy1} A dense $20\times20\times10$ k-mesh was used to check the dynamical stability. To go beyond quasi-harmonic approximation, Phono3py\cite{phono3py} was used to calculate lattice thermal conductivity which incorporates third-order inter-atomic force constants. A $2\times2\times2$ supercell was used within the small displacement method (with a displacement criteria of 0.03 {\AA}) which generated 4738 supercells. Phonon-lifetime was calculated using the tetrahedron method involving a dense $24\times24\times12$ q-mesh in a temperature range of 300 to 900 K. To analyze the nature of chemical bonding between atoms, the crystal-orbital Hamilton-population (COHP) was calculated using LOBSTER code\cite{LOBSTER1,LOBSTER2,LOBSTER3} with mBJ exchange-correlation functional. To calculate the electronic transport properties, Ab initio Scattering and Transport (AMSET)\cite{AMSET} code was used which solves the Boltzmann transport equation considering the energy and temperature-dependent relaxation time (see supplementary material (SM)\cite{suppl1} for more details on formalism). AMSET captures various kinds of scattering including Acoustic Deformation Potential (ADP), Polar Optical Phonon (POP), Piezoelectric (PIE) and Ionized Impurity (IMP) scattering to determine the variable relaxation time.
\begin{figure}[t]
	\centering
	\includegraphics[scale=0.27]{ 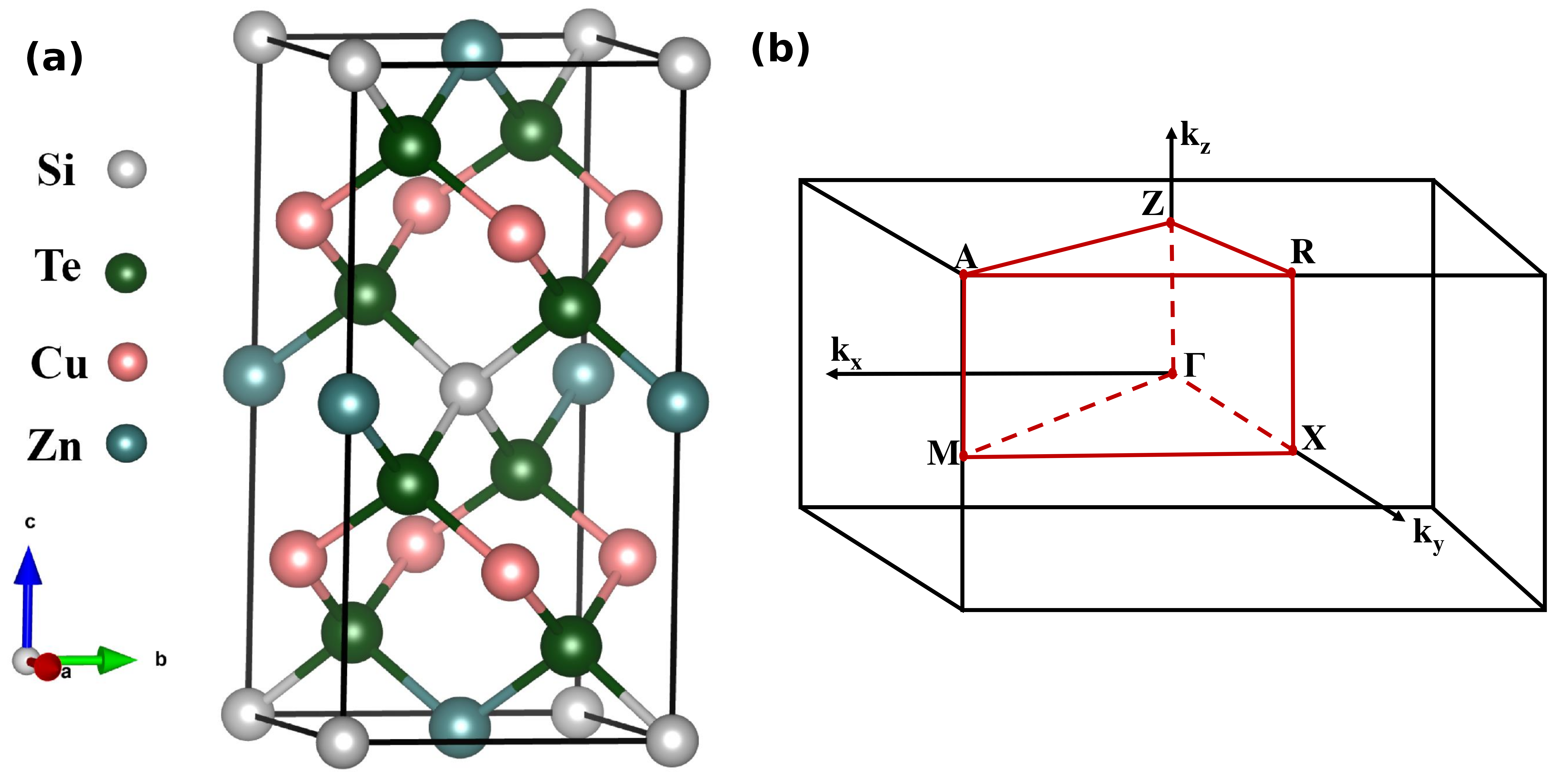}
	\caption{ For bulk Cu$_2$ZnSiTe$_4$, (a) crystal structure with space group I$\bar{4}$2m and (b) corresponding Brillouin zone }
	\label{Crystal Structure}
\end{figure}
\section{Results and Discussions}
\subsection{Crystal Structure}
Cu$_2$ZnSiTe$_4$ is experimentally reported\cite{RamanCu2Te4} to crystallize in tetragonal symmetry with space group I$\bar{4}$2m ($\#$121) (D$_2d$ point group) and lattice parameter a = b = 5.961 \AA \hspace{1mm}  and c = 11.788 \AA \hspace{1mm} at room temperature\cite{RamanCu2Te4}. Figure \ref{Crystal Structure}(a,b) shows the crystal structure and its corresponding Brillouin zone with the high symmetry points.
\begin{figure}[t]
\centering
\includegraphics[width=3.4in]{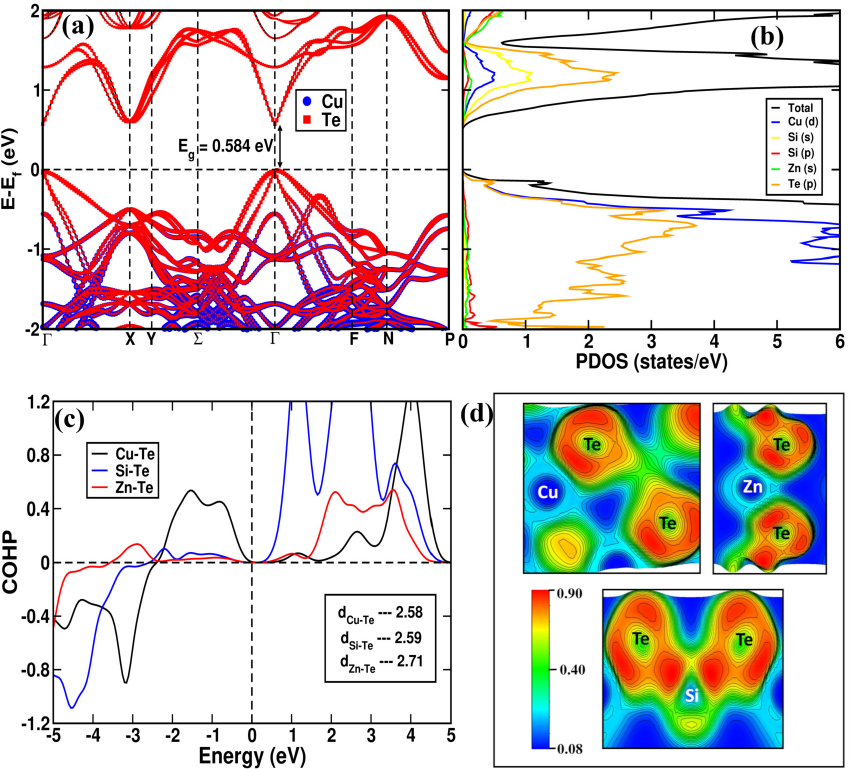}
\caption{For Cu$_2$ZnSiTe$_4$, (a) Cu- and Te-projected electronic band structure, (b) orbital projected density of states (PDOS) using mBJ exchange-correlation functional including spin orbit coupling. (c) calculated crystal orbital Hamilton population (COHP) and (d) Electron Localization function (ELF) for Cu-Te, Si-Te and Zn-Te pairs.  }
\label{bands_electron}
\end{figure}
 Our theoretically optimized lattice parameters are a = b = 6.05 \AA \hspace{1mm} and c = 11.99 \AA, which agrees fairly well with the experiment values with a mismatch of $\sim$ 1.4 \% in a-axis and $\sim$ 1.7 \% in c-axis. It has two formula units leading to 16 atoms in the unit cell. The structure can be derived starting from a zinc-blend (ZnS) type-structure by making a supercell along the z-direction and replacing zinc with three different cations (Cu, Zn and Si) and Sulphur (S) with tellurium (Te). This makes these quaternary chalcogenides a naturally distorted supercell structure. Every Te (anion) atom is located in a tetrahedral void formed by other cations.  Due to the different atomic radii of Cu, Zn and Si, anion (Te) makes three different bonds i.e. d$_{Cu-Te} = 2.578$  \AA \hspace{1mm} (Cu-Te), d$_{Si-Te} = 2.586$  \AA \hspace{1mm} and d$_{Zn-Te} = 2.706$  \AA \hspace{1mm} in the same tetrahedron. In other words, Te forms a locally distorted environment which, in turn, helps in reducing the thermal conductivity.
\subsection{Electronic structure}
\begin{figure}[t]
\centering
\includegraphics[scale=1.2]{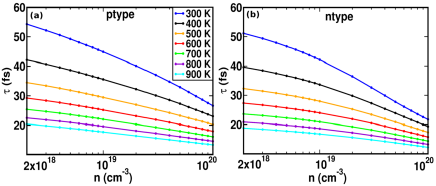}
\caption{For Cu$_2$ZnSiTe$_4$, carrier concentration (n) dependence of relaxation time for (a) holes (p-type) and (b) electrons (n-type) at different temperatures. }
\label{relax_time}
\end{figure}

\begin{figure*}[t!]
\centering
\includegraphics[scale=0.75]{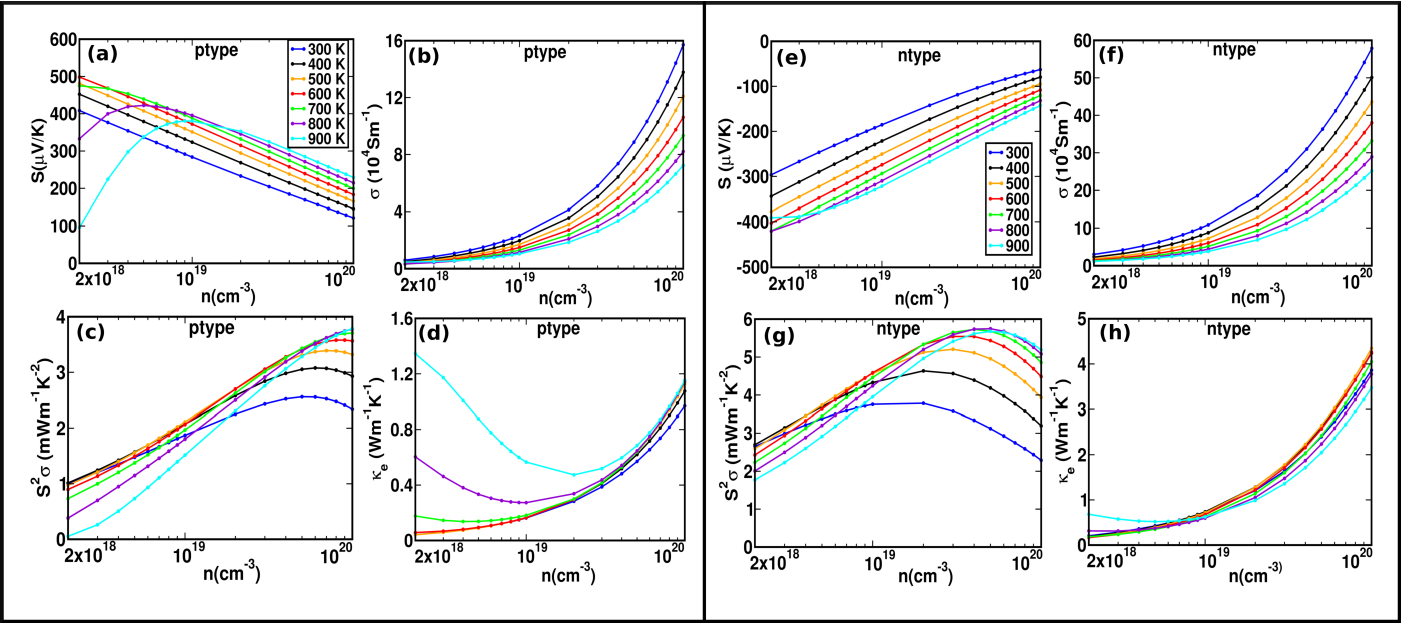}
\caption{For Cu$_2$ZnSiTe$_4$, carrier concentration (n) dependence of (a,e) Seebeck coefficient (S) (b,f) electrical conductivity ($\sigma$) (c,g) power-factor (S$^2\sigma$) and (d,h) electronic thermal conductivity ($\kappa_e$) for p-type (left) and n-type (right) conduction respectively at different temperatures (T). }
\label{TE_properties}
\end{figure*}
Figure \ref{bands_electron}(a,b) shows the electronic band structure and atom/orbital projected density of states (PDOS) of Cu$_2$ZnSiTe$_4$ including the effect of spin orbit coupling. In the absence of SOC, PBE functional yields a band gap of 0.17 eV which gets corrected to 0.74 eV when meta-GGA mBJ functional is used. SOC shows an appreciable effect on the band structure topology (see comparison of band structure with and without SOC in Fig. SI of SM\cite{suppl1}), including a reduction in the band gap value from 0.74 eV to 0.584 eV. Below VBM, bands are relatively flatter compare to CBM implying the possibility of holes becoming the dominant charge carriers in TE transport, which indeed gets reflected in our simulated Seebeck coefficient values (see Sec. D below). Cu and Te atoms together provide the major contribution around valence band maxima (VBM), while Te atoms contribute maximum near conduction band minima (CBM), as is illustrated in the PDOS plot in Fig. \ref{bands_electron}(b). Further analysis of the PDOS reveals that states below E$_{\text{F}}$ are primarily made up of Cu(d) and Te(p) orbitals, which overlap in the small energy range up to 0.6 eV.   Nevertheless, the region above E$_{\text{F}}$ may be separated into two sections: the first includes dominant Te p-orbital, with a minor contribution from Si s-orbital, and the second includes a region between 1.5 - 2.0 eV, which is composed of Zn s-orbital, Te, and Si p-orbital. In other words, this structure can be considered to comprise of two functional units (i) a conduction unit Cu$_2$Te$_4$ which dominates at/around E$_{\text{F}}$ and (ii) an insulating unit ZnSiTe$_4$ away from E$_{\text{F}}$.

To assess the bonding nature between different atoms, we have simulated the band energy in terms of orbital pair contribution using crystal orbital Hamilton population (COHP) for Cu-Te, Zn-Te and Si-Te bonding interactions. Figure \ref{bands_electron}(c) shows the COHP vs. energy highlighting the energy-dependent bonding (negative) and anti-bonding (positive) character of different cations and anions in Cu$_2$ZnSiTe$_4$. Since there are no states at Fermi level (E$_{\text{F}}$), it signifies the stable bonding interaction and semiconducting nature of the compound\cite{PbTe}. The interaction below the E$_{\text{F}}$ is dominated by Cu-Te pair. This pair shows anti-bonding states due to the localization of charge, as reflected in the charge density plots in Fig. \ref{bands_electron}(c). Another plausible reason for anti-bonding states near E$_{\text{F}}$ can be the locally distorted structure, as discussed above in the bond length analysis. To further analyze the bonding nature, we have plotted the electron localization function (ELF)\cite{ELF} in Fig. \ref{bands_electron}(d). A value of ELF$\sim$ 1 indicates a strong localization nature, whereas an electron gas is represented by an ELF of $\sim$ 0.5. Since the tetrahedral void anion (Te) is linked to various cations (Cu, Zn, and Si),  ELF plots are shown for all three pairs Cu-Te, Zn-Te, and Si-Te. Similar electron localization around the Te atom is observed in case of Cu-Te and Zn-Te pairs, indicating an ionic bonding. On the other hand, Si-Te pair indicates a covalent-like bonding characteristic.

\begin{figure}[t]
\centering
\includegraphics[width=3.2in]{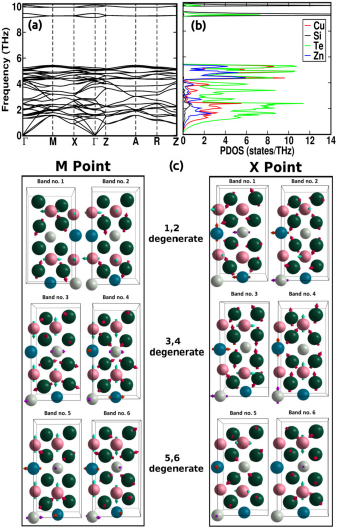}
\caption{For Cu$_2$ZnSiTe$_4$, (a) phonon dispersion (b) atom/orbital projected phonon density of states (PDOS) and (c) different phonon modes at M and X high symmetry points with band degeneracy. }
\label{bands_phonon}
\end{figure}

\subsection{Electronic Transport}

A temperature and energy dependent variable relaxation time approach is used to simulate the transport coefficients (S, $\sigma$ and $\kappa_e$). This makes these quantities more accurate/reliable as compared to those simulated using constant relaxation time approximation (CRTA)\cite{CRT5,CRT3,CRT8}. In doped semiconductors, scattering mediated by ADP, POP and IMP play very crucial role. The variable relaxation time can be calculated using Fermi golden rule as,
\begin{equation}
\tau_{nk \rightarrow mk+q}^{-1} = \frac{2\pi}{\hbar}|g_{nm}(k,q)|^2\delta(\epsilon_{nk}-\epsilon_{mk+q})
\end{equation}
where $g_{nm}(k,q)$ is the matrix element for scattering from state  $|nk\rangle$ to state $|mk+q\rangle$ and $\epsilon_{nk}$ is the energy of the state $|nk\rangle$. Further details on  the scattering specific  matrix elements are discussed in Sec. II of SM\cite{suppl1}.
Figure \ref{relax_time} shows the carrier concentration (n) dependence of relaxation time ($\tau$) at different temperatures for n- and p-type conduction. This relaxation time is calculated by incorporating the effect of all the four scatterings, i.e. ADP, POP, PIE and IMP. Relaxtion times for each of the individual scatterings are shown in Fig. SII of SM\cite{suppl1}. Total  $\tau$ for p-type conduction varies between 55 and 13 fs in the entire concentration and temperature range considered here, while the same for n-type conduction varies between 52 and  10 fs. Increasing temperature leads to a reduction in relaxation time due to the enhanced phonon scattering. Doping can be an efficient strategy to improve electrical conductivity. For example, excess Cu doping in Cu$_2$MnSnSe$_4$ is reported to increase the value of conductivity by 1 order of magnitude from $10^3$ to $10^4$ (S/m). Similarly, for n-type quaternary compound AgPbBiSe$_3$\cite{AgPbBiSe1}, conductivity value increases from 5$\times$$10^3$ to 2.5$\times$$10^4$ (S/m) with chlorine (Cl) doping. As such, for the power-factor (S$^2\sigma$) to enhance, the electrical conductivity should increase to such an extent that it can overcome the reduction in S, as a result of doping. To explore this, we have done calculations in the doping concentration range of 10$^{18}$-10$^{20}$ cm$^{-3}$. The effect of doping is incorporated within the rigid band approximation (RBA) which considers the topology of bands to remain intact as that of the parent system and vary the Fermi level in the band structure\cite{RBA}. Seebeck coefficient values are found to be higher in p-type as compared to n-type conduction due to the flat valence bands. At high enough temperature, minority carriers may be generated which can affect the conduction mechanism, often known as bipolar thermal conduction. It is dominant at high temperatures and low carrier concentrations. Figure \ref{TE_properties} shows the carrier concentration (n) dependence of S, $\sigma$, $S^2\sigma$ and electronic thermal conductivity ($\kappa_e$). Clearly, S decreases as n increases while $\sigma$ follows the opposite trend which eventually helps to enhance power-factor ($S^2\sigma$). The latter approaches a maximum value at an optimal carrier concentration and then starts to decrease. Bipolar effect is visible at high temperatures and low n, which makes electronic thermal conductivity ($\kappa_e$) increase in this regime. Clearly, peaks of TE coefficients occur at different carrier concentrations for n- and p-type conduction, which is due to the different nature of band topology resulting in different scattering rates.\par
The values of power-factor are comparable (or better) to some of the promising TE materials in this family Cu$_2$ABX$_4$. S$^2\sigma$ for n-type conduction at 900 K is 3.95 mWm$^{-1}$K$^{-2}$ at the optimal carrier concentration of $1\times10^{19}$ cm$^{-3}$, which can be compared with S$^2\sigma$= 0.5 mWm$^{-1}$K$^{-2}$ of Cl-doped AgPbBiSe$_3$\cite{AgPbBiSe1}. For p-type conduction,  S$^2\sigma$ shows a peak value of 3.063 mWm$^{-1}$K$^{-2}$ at the optimal carrier concentration of 4$\times10^{19}$ cm$^{-3}$, which can be compared with the Cu doped (excess) Cu$_2$MnSnSe$_4$ with the highest power-factor of 0.71 mWm$^{-1}$K$^{-2}$. Enhancement in electrical conductivity is predominantly responsible for such high values of power-factor.
\subsection{Phonon properties}
Figure \ref{bands_phonon}(a) shows the phonon dispersion of Cu$_2$ZnSiTe$_4$. The low frequency acoustic phonons acquire small group velocity responsible for low lattice thermal conductivity. The group velocity for three acoustic branches LA, TA and TA$'$ are 3306.96, 1717.18 and 2729.70  m/s respectively.

Mixing of low energy optical phonons with acoustic phonons enhances the phonon scattering leading to suppression of thermal transport. Te atom which resides in the tetrahedral void (surrounded by other atoms) vibrates in the free space of the void and scatters the phonon more. Nonetheless, the Te atom contributes dominantly to the low acoustic modes, as illustrated by the atom projected density of states in Fig. \ref{bands_phonon}(b).   The lowest optical branch which gets entangled with the acoustic branches lie at $\sim$ 2 THz.  Te atoms show appreciable contribution in this region as well, along with the Cu atoms.  In the frequency range 4-6 THz, states of Te-atoms hybridize strongly with those of Cu and Zn, resulting in higher density of states and hence enhancement of phonon scattering. In order to show the nature of phonon propagation, we have plotted the eigen modes of different bands at two high symmetry points `M' and `X' in Fig. \ref{bands_phonon}(c). It clearly shows the degeneracy of (1,2), (3,4) and (5,6) pairs of bands at these two high symmetry points. The first two degenerate modes are acoustic, 3rd and 4th bands are acoustic and optical respectively while the last two are optical modes. Appreciable mixing of acoustic and low lying optical phonon modes is responsible for the low lattice thermal conductivity, as discussed below.\par
\begin{figure}[t]
\centering
\includegraphics[width=3.49in]{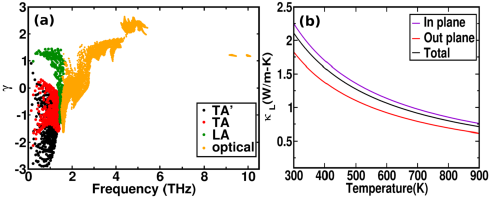}
\caption{For Cu$_2$ZnSiTe$_4$, (a) Gruneisen parameter ($\gamma$) vs. phonon frequency and  (b) in-plane and out-of-plane component of lattice thermal conductivity ($\kappa_L$) vs. temperature. }
\label{lt_thermal}
\end{figure}

Another important physical parameter that plays a crucial role in evaluating the lattice thermal conductivity is the Gruneisen parameter ($\gamma = -\frac{\partial \log \omega_i}{\partial V_i}  $) which is also a measure of the degree of anharmonicity. Figure \ref{lt_thermal}(a) shows the phonon frequency dependence of $\gamma$ for different phonon modes. One can notice that $\gamma$ is mostly positive in the higher frequency range, and shows negative values predominantly at low frequencies for acoustic modes. This implies that frequency will increase with the increase in volume.  This might be due to the unique structure framework in the present case which consists of numerous cation-anion bonds, with the anion at the centre of the tetrahedral void (corner shared by cations). At low frequencies, transverse displacement of the anion (Te) atom causes a contraction of other cationic atoms towards each other in the void.\cite{NTE_review} Figure \ref{lt_thermal}(b) shows the temperature dependence of calculated lattice thermal conductivity ($\kappa_L$).  These calculations are done by incorporating inter-atomic force constants simulated up to third order which is expected to provide higher accuracy. Simulated $\kappa_L$ is found to decrease from 2.1 to 0.7 Wm$^{-1}$K$^{-1}$ with increasing temperature from 300 to 900 K.

\begin{figure}[t]
 \centering
 \includegraphics[width=3.4in,height=2.2in]{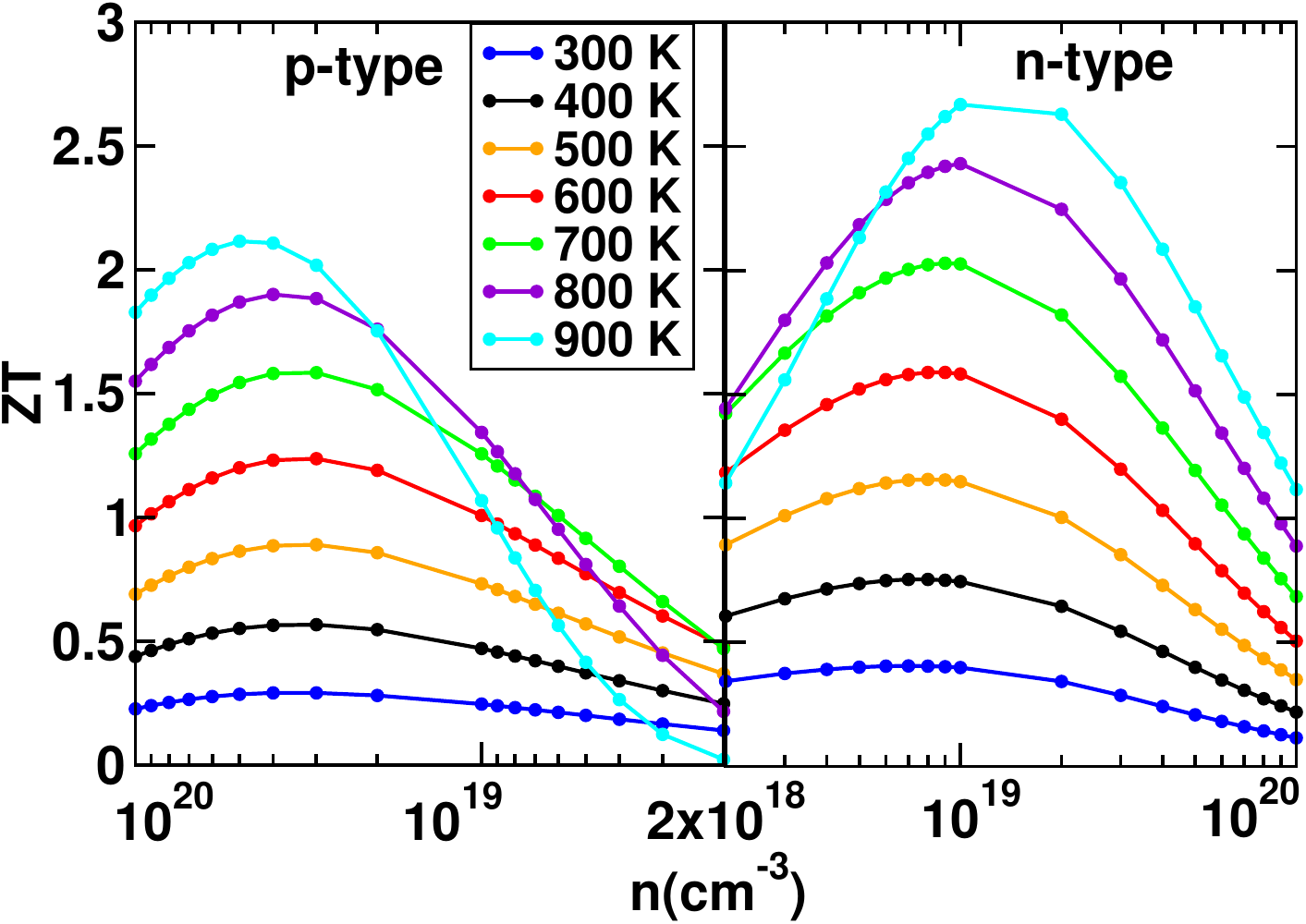}
 \caption{For Cu$_2$ZnSiTe$_4$, carrier concentration (n) dependence of TE figure of merit (ZT) for p-type (left) and n-type (right) conduction respectively at different temperatures (T). }
 \label{merit}
 \end{figure}

The $\kappa_L$ values for few experimentally reported compounds belonging to this family such as Cu$_{2}$(Zn/Co/Mn/Fe)SnSe$_{4}$\cite{Co_Mn3,ZnSn1}, Cu$_{2}$HgSnTe$_{4}$\cite{HgSn1} and Cu$_{2}$(Cd$/$Zn)GeSe$_{4}$\cite{Cd_Ge,Zn_Ge}  varies in the range 2.8 to 1.3 Wm$^{-1}$K$^{-1}$ at room temperature while reduces to the range 1.9 to 0.5 Wm$^{-1}$K$^{-1}$ at higher temperatures (600 to 800 K). Our calculated lattice thermal conductivity for Cu$_2$ZnSiTe$_4$ lie in a similar range. The theoretically reported $\kappa_l$ values for few of the compounds belonging to the same family e.g, Cu$_2$CdSnTe$_4$\cite{p3py2} and CuZn$_2$InTe$_4$\cite{p3py1} using Phono3py lie in the range 2.0 to 2.1 Wm$^{-1}$K$^{-1}$  at room temperature.
After evaluating the electronic and lattice thermal transport properties, we have estimated the ZT value for Cu$_2$ZnSiTe$_4$. Figure~\ref{merit} depicts the carrier concentration dependence of ZT values at different temperatures. Our calculation shows reasonably high ZT values of 2.11 for p-type and 2.67 for n-type at 1$\times10^{19}$ cm$^{-3}$ and 5$\times10^{19}$ cm$^{-3}$ concentrations respectively at 900 K.
\section{Conclusions}
In summary, we report a detailed study of the electronic and thermal transport properties of a quaternary chalcogenide Cu$_2$ZnSiTe$_4$ using first-principles calculation. This compound consists of two distinct structural units (Cu$_2$Te$_4$ and ZnSiTe$_4$) one of which is favourable for high electrical conduction while the other for thermal transport. In other words, it gives a flexibility to tune both types of conductions which is highly required for TE materials. The low band gap ($\sim$0.58 eV) along with large spin orbit coupling effects (due to Te-atoms) help to achieve reasonably high power factor, as large as $\sim$ 3.95 and 3.06 mWm$^{-1}$K$^{-2}$ for n- and p-type conductions respectively. Distorted structural framework (involving cation-anion bonding, with anion at the centre of the tetrahedra void) helps to entangle the acoustic and low energy optical phonon modes, which in turn is responsible for reduced lattice thermal conductivity (lying in the range $\sim$ 2.0 to 0.7 Wm$^{-1}$K$^{-1}$). Large power-factor and moderate $\kappa_L$ values collectively make this compound a potential candidate for TE applications with TE figure of merit (ZT)  $\sim$ 2.11 (p-type) and 2.67 (n-type) at high temperature. We strongly believe that the present work can pave a path for future experimental studies on Cu$_2$ZnSiTe$_4$ for TE applications.

\section{Acknowledgements:} HS and BS acknowledge the support of HPC facility (spacetime-2) at IIT Bombay. HS also acknowledges the computational support of the Thematic Unit of Excellence on Computational Materials Science, funded by the Nano Mission of the Department of Science and Technology at S N Bose National Centre for Basic Science, Kolkata. TSD acknowledges a J.C.Bose National Fellowship (grant no. JCB/2020/000004) for funding.

\bibliographystyle{unsrtnat}
\bibliography{main.bib}
\end{document}